\documentclass[a4paper,12pt]{article}
\makeatletter

\@addtoreset{equation}{section}
\makeatother

\usepackage[nosort]{cite}

\begin{document}

\pagestyle{plain}
\setcounter{page}{1}


\begin{titlepage}

\rightline{\tt hep-th/0301176}

\vskip .1 cm
\rightline{\small{\tt TU-679}}

\begin{center}

\vskip 2 cm
{\Large\bfseries Noncommutative $U(1)$ Instantons in}
\vskip 0.5cm
{\Large\bfseries Eight Dimensional Yang-Mills Theory} 

\vskip 2cm
{\large Yoshiki Hiraoka}

\vskip 1.2cm
{\it Department of Physics, Tohoku University}
{\it Sendai 980-8578, JAPAN}

\vskip .7cm
{\tt hiraoka@tuhep.phys.tohoku.ac.jp}

\vskip 1.5cm


{\bf Abstract}
\end{center}

\noindent

We study the noncommutative version of the extended ADHM construction 
in the eight dimensional $U(1)$ Yang-Mills theory. 
This construction gives rise to the solutions of the BPS equations 
in the Yang-Mills theory, 
and these solutions preserve at least 3/16 of supersymmetries.  
In a wide subspace of the extended ADHM data, 
we show that the integer $k$ which appears in the extended ADHM construction 
should be interpreted as the $D4$-brane charge 
rather than the $D0$-brane charge  
by explicitly calculating the topological charges 
in the case that the noncommutativity parameter is anti-self-dual. 
We also find the relationship with the solution generating technique 
and show that the integer $k$ can be interpreted 
as the charge of the $D0$-brane bound to the $D8$-brane with the $B$-field 
in the case that the noncommutativity parameter is self-dual.

\end{titlepage}

\newpage


\section{Introduction}


Noncommutative geometry has played an important role 
in the study of string/M-theory~\cite{cds}.
In particular, $D$-branes with a constant NS $B$-field are 
of interest in the context of understanding 
the non-perturbative aspects of string theory.
The effective world-volume field theory on $D$-branes with a $B$-field 
turns out to be the noncommutative Yang-Mills theory~\cite{sw}, 
which has an interesting feature 
that the singularity of the instanton moduli space 
is naturally resolved~\cite{ns}.

Four dimensional $U(N)$, $k$ instantons are realized as $k$ $D0$-branes 
within $N$ $D4$-branes in type IIA string theory.
When we turn on a constant $B$-field  
which preserves 1/4 of supersymmetries,  
the moduli space of the noncommutative instantons is resolved 
and the $D0$-branes cannot escape from the $D4$-branes.
From the viewpoint of the $D0$-brane theory, 
the Higgs branch of the moduli space coincides with the moduli space 
of the noncommutative instantons 
and the $B$-field corresponds to the Fayet-Iliopoulos parameters.

The instanton solutions of the Yang-Mills theory are constructed 
by the well-known ADHM method.  
There is the one-to-one correspondence 
between the moduli space of the instantons and that of the ADHM data 
in the commutative case~\cite{adhm, cg}. 
On the other hand, 
most of the noncommutative instantons in four dimensions 
have been obtained by modifying the ADHM construction. 
See e.g.~\cite{ns, nekrasov, furuuchi, furuuchi2, furuuchi3, 
agms, kly, ckt, hamanaka, popov2, sako1, sako2, sako3, 
meron, dress, ly, lty, kly2, ly2} 
and references therein. 
In particular, it has been proven that 
the instanton number is generally an integer 
in the noncommutative $U(N)$ gauge theory by Sako~\cite{sako3}.

It is also of interest to generalize the above system 
to higher dimensions in the context of both $D$-brane dynamics 
and the world-volume theories. 
The systems of the $D0$-$D6$ and the $D0$-$D8$ with the $B$-field 
have been investigated by several authors 
~\cite{ohtan, cimm, park, witten, sato, fio, ohta, hio, 
kly3, pt, hiraoka, hiraoka2, blp, valtancoli}.  
Especially we consider the system of the $D0$-brane 
and the $D8$-brane with the $B$-field. 
These studies reduce to finding the solutions of 
the higher dimensional analogue 
of the ``self-duality'' equations 
which are the first order linear relations 
amongst the components of the field strength~\cite{cdfn, ward, hull, blp}.    
It has been shown that there are many kinds of the BPS equations
which preserve $1/16, 2/16$, $\cdots$, $6/16$ of supersymmetries 
and these equations are related to the subgroup $SO(7)$, $SO(6)$, $\cdots$, 
$SO(2)$ of the eight dimensional rotation group $SO(8)$.

In this paper, we focus on the case 
that is related to the $SO(5)=Sp(2)$ symmetry. 
In this case the configuration of the gauge field 
preserves at least 3/16 of supersymmetries. 
It is known that there is the extended ADHM construction 
which gives rise to the solutions of the 3/16 BPS equations 
in eight dimensions~\cite{cgk}. 
We consider the noncommutative $U(1)$ gauge theory 
and study the noncommutative version 
of this extended ADHM construction~\cite{ohta}. 
It is worth constructing the simple solutions explicitly and 
investigating their properties such as the topological charges
since a little thing about the noncommutative version 
of the extended ADHM construction is known until now. 
This subject has been studied in some references~\cite{ohta, hio, hiraoka}.

This paper is organized as follows. 
In section 2, we review the BPS equations and 
the extended ADHM construction in eight dimensions.
In section 3, we briefly review the Yang-Mills theory 
on the noncommutative space. 
As in the four dimensional case, it is an important difference  
whether the noncommutativity parameter is anti-self-dual or self-dual. 
In section 4 we consider the case 
that the noncommutativity parameter is anti-self-dual. 
In a wide subspace of the extended ADHM data, 
we show that the integer $k$ which appears in the extended ADHM construction 
should be interpreted as the $D4$-brane charge 
rather than the $D0$-brane charge
by explicitly calculating the topological charges. 
In section 5, we consider the case 
that the noncommutativity parameter is self-dual. 
We find the relationship with the solution generating technique 
and show that the integer $k$ can be interpreted 
as the charge of the $D0$-brane bound to the $D8$-brane with the $B$-field. 
The final section is devoted to the conclusion.


\section{Extended ADHM construction of eight dimensional instantons}


In this section, we review the BPS equations 
and the extended ADHM construction of the instantons in eight dimensions. 
The instantons in higher dimensions are defined 
as the solutions of the BPS equations. 
This definition is the natural generalization 
of the four dimensional $U(N)$, $k$ instantons 
which are constructed by the ADHM construction 
with the gauge group $U(N)$ and the ADHM parameter $k$. 
These instantons have the $D$-brane interpretation 
as the bound states of $k$ $D$0-branes and $N$ $D$4-branes. 

Therefore, in the following, 
we consider the extended ADHM construction 
with the gauge group $U(N)$ and the extended ADHM parameter $k$ 
since these situations are expected to correspond to 
the systems of $k$ $D0$-branes and $N$ $D8$-branes.


\subsection{BPS equations in eight dimensions}


In this subsection, we briefly review the BPS equations 
in the eight dimensional Yang-Mills theory. 
These equations were studied in Ref.~\cite{cdfn, ward, hull}, 
and systematically classified by the authors of~\cite{blp}. 
The BPS equations are the higher dimensional analogue 
of the ``self-duality'' equations, 
which are the linear relations amongst the components of the field strength 
\begin{equation}
 \frac{1}{2}T_{\mu\nu\rho\sigma}F_{\rho\sigma}=F_{\mu\nu}\,,
 \quad (\mu,\,\nu,\,\rho,\,\sigma =1,\,\cdots,\, 8)\,,\label{eq:2.1}
\end{equation}
with the constant 4-form tensor $T_{\mu\nu\rho\sigma}$\,. 
These equations are the natural generalizations 
of the four dimensional self-duality equations:
\begin{equation}
  \frac{1}{2}\epsilon_{abcd}F_{cd}=F_{ab}\,,
 \quad (a,\,b,\,c,\,d =1,\,\cdots ,\,4)\,. \label{eq:2.2}
\end{equation}

When the equations (\ref{eq:2.1}) hold, 
the equations of motion, $D_{\mu}F_{\mu\nu}=0$,  
are automatically satisfied due to the Jacobi identity, 
and the lower bound of the action is reached.  
This bound is obtained as in the four dimensional case by the identity
\begin{equation}
 -\frac{1}{4}F_{\mu\nu}F_{\mu\nu} 
 = -\frac{1}{16}\left( F_{\mu\nu}-\frac{1}{2}T_{\mu\nu\rho\sigma}
F_{\rho\sigma}\right)^2
-\frac{1}{8} T_{\mu\nu\rho\sigma}F_{\mu\nu}F_{\rho\sigma}\nonumber\,,
\label{eq:2.3}
\end{equation}
where the gauge field is taken to be anti-hermitian. 
This identity was shown by the authors of~\cite{blp}.

It was also shown in Ref.~\cite{blp} 
that there are many kinds of the BPS equations 
which preserve $1/16, 2/16,$ $\cdots,$ $6/16$ of supersymmetries
and these equations are related to the subgroup 
$SO(7)$, $SO(6)$ $\cdots$, $SO(2)$ 
of the eight dimensional rotation group $SO(8)$. 
Especially in this paper 
we concentrate on the case that is related to the $SO(5)=Sp(2)$ symmetry. 
In this case the configuration of the gauge field preserves 
at least $3/16$ of supersymmetries.


\subsection{Extended ADHM construction on $\mathbf{R}^8$}


The ADHM construction is a powerful tool to construct 
the Yang-Mills instantons in four dimensions~\cite{adhm, cg}. 
Especially, it is well-known that the instanton moduli space and the 
ADHM moduli space completely coincide in four dimensions. 
It is also known that there exists the extended ADHM construction 
which gives rise to solutions of the 3/16 BPS equations in eight dimensions. 
This extended ADHM construction was investigated 
by several authors~\cite{cgk, ohta, hio, pt, hiraoka, hiraoka2}.

In this subsection, we review this extended ADHM construction of the 
eight dimensional ``self-dual'' instantons associated with the $Sp(2)$ 
group given in~\cite{ward, cgk}.
This construction of the instantons is the slight extension of 
the four dimensional ADHM construction.
When we take $\mathbf{B}=0\,,$ or $\mathbf{B}'=0$ which are defined 
in the following, 
we reproduce the four dimensional ADHM construction.

In order to treat the eight dimensional space, 
it is useful to regard eight coordinates of the $\mathbf{R}^8$ 
as two quaternionic coordinates: 
\begin{equation}
\mathbf{x}=\sum_{\mu =1}^8 \tilde{\sigma}_{\mu}x^{\mu}=
\left( \begin{array}{@{\,}cc@{\,}}
   z_2 & z_1\\
  -\bar{z}_1  &  \bar{z}_2
  \end{array}  \right)\,,\quad 
\mathbf{x}'=\sum_{\mu =1}^8 \tilde{\sigma}'_{\mu}x^{\mu}=
\left( \begin{array}{@{\,}cc@{\,}}
   z_4 & z_3\\
  -\bar{z}_3  &  \bar{z}_4
  \end{array}  \right)\,. \label{eq:2.4}
\end{equation}
Here we defined the eight vector matrices
using the Pauli matrices $\tau_i$ $(i=1,\,2,\,3)$ by
\begin{eqnarray}
\tilde{\sigma}_{\mu}&=& (\,i\tau_1,\,0,\,\,\,i\tau_2,\,0,\,\,\,
i\tau_3,\,0,\,\,\,\textbf{1}_2,\,0\,\,)\,,\label{eq:2.5}\\
\tilde{\sigma}'_{\mu}&=&(\,0,\,\,\,i\tau_1,\,0,\,\,\,i\tau_2,\,
0,\,\,\,i\tau_3,\,0,\,\,\textbf{1}_2\,)\,,\nonumber
\end{eqnarray}
and the four complex coordinates by
\begin{equation}
 z_1=x^3+ix^1\,,\quad z_2=x^7+ix^5\,,\quad 
z_3=x^4+ix^2\,,\quad z_4=x^8+ix^6\,.\label{eq:2.6}
\end{equation}

The four dimensional ADHM construction gives rise to the instantons 
through the zero mode of a zero dimensional massless Dirac-like operator. 
This construction can be easily extended to the eight dimensional case. 
At first we define the Dirac-like operator 
\begin{equation}
D_z = \textbf{A}+\overrightarrow{\mathbf{B}}\cdot \overrightarrow
{\mathbf{X}}\,,\label{eq:2.7}
\end{equation} 
using the $(N+2k)\times 2k$ matrices $\mathbf{A}\,,\,\mathbf{B}$ 
and $\mathbf{B}'$. 
Here we also define $\overrightarrow{\mathbf{B}}=(\mathbf{B},\mathbf{B}')$ 
and $\overrightarrow{\mathbf{X}}=(\mathbf{x},\mathbf{x}')$\,. 
Then we can construct the $U(N)$ gauge field as
\begin{equation}
 A_{\mu}=\psi^{\dagger}\partial_{\mu}\psi\,,\label{eq:2.8}
\end{equation}
where the $(N+2k)\times N$ matrix $\psi$ is 
the solution of the following Dirac-like equation
\begin{equation} 
D_z^{\dagger}\psi=0\,,\label{eq:2.9}
\end{equation} 
and is normalized as 
$\psi^{\dagger}\psi =\mathbf{1}_{N\times N}$\,.

We can see that the gauge field (\ref{eq:2.8}) gives 
the ``self-dual'' field strength as
\begin{eqnarray}
F_{\mu\nu}&=& 2\psi^{\dagger}\left( \partial_{\left[ \mu\right. } D_z
\frac{1}{ D^{\dagger}_zD_z}  \partial_{\left. \nu \right]  }D^{\dagger}_z
\right)\psi\nonumber\\
&=& 2\psi^{\dagger}\overrightarrow{\mathbf{B}}\,
\overline{N}_{\mu\nu}\frac{1}{ D^{\dagger}_zD_z}
\overrightarrow{\mathbf{B}}^{\dagger}\psi\,,\label{eq:2.10}
\end{eqnarray}
where we used the completeness relation
\begin{equation}
\textbf{1}_{N+2k}=\psi\psi^{\dagger}
+D_z\frac{1}{ D^{\dagger}_zD_z}D^{\dagger}_z\,,\label{eq:2.11}
\end{equation}
and defined the ``self-dual'' tensor $\overline{N}_{\mu\nu}$\,.
This tensor is explicitly written as
\begin{equation}
\overline{N}_{\mu\nu}=\frac{1}{2}(\Sigma_{\mu}\Sigma_{\nu}^{\dagger}-
\Sigma_{\nu}\Sigma_{\mu}^{\dagger})\,,\label{eq:2.12}
\end{equation}
by using the following quantity,
\begin{equation}
\Sigma_{\mu}\equiv \partial_{\mu}\overrightarrow{\mathbf{X}}=
\left( \begin{array}{@{\,}c@{\,}}
   \tilde{\sigma}_{\mu} \\
  \tilde{\sigma}'_{\mu}
  \end{array}  \right)\,,
\end{equation}
and satisfies the ``self-duality'' equation, 
\begin{equation}
  \frac{1}{2}T_{\mu\nu\rho\sigma}\overline{N}_{\rho\sigma}=
\overline{N}_{\mu\nu}\,,\label{eq:2.14}
\end{equation}
where $T_{\mu\nu\rho\sigma}$ is the $Sp(2)$ invariant tensor.

The identity (\ref{eq:2.14}) is surely identical 
to the ``self-duality'' equation (\ref{eq:2.1})\,.
Therefore we can automatically construct 
the solutions of the 3/16 BPS equations 
through the extended ADHM construction given above. 
In our choice of the complex coordinates (\ref{eq:2.6}), 
we can write down the general form of the field strength
which satisfies the ``self-duality'' equation (\ref{eq:2.14}) as
\begin{eqnarray}
F&=& F_{z_1\bar{z}_1}\left( dz_1\wedge d\bar{z}_1 +dz_2\wedge d\bar{z}_2
\right)+F_{z_3\bar{z}_3} \left(dz_3\wedge d\bar{z}_3 +dz_4\wedge d\bar{z}_4 
\right)\nonumber\\
& & \quad +F_{z_1\bar{z}_3} \left( dz_1\wedge d\bar{z}_3 
+dz_2\wedge d\bar{z}_4\right) +F_{\bar{z}_1z_3}\left(  
d\bar{z}_1\wedge dz_3 +d\bar{z}_2\wedge dz_4\right)\nonumber\\
& & \quad  +F_{z_1z_4}\left( dz_1\wedge dz_4 +dz_3\wedge dz_2
\right) +F_{\bar{z}_1 \bar{z}_4}\left( d\bar{z}_1\wedge d\bar{z}_4 
+d\bar{z}_3\wedge d\bar{z}_2\right)\nonumber\\
& & \quad +F_{z_1z_2}dz_1\wedge dz_2 +F_{\bar{z}_1\bar{z}_2}d\bar{z}_1\wedge 
d\bar{z}_2 \nonumber\\
& & \quad +F_{z_3z_4}dz_3\wedge dz_4 
+F_{\bar{z}_3\bar{z}_4}d\bar{z}_3\wedge d\bar{z}_4 \,.\label{eq:2.15}
\end{eqnarray}


\subsection{Extended ADHM equations}


It is crucial that the $D^{\dagger}_zD_z$ is invertible and commutes 
with $\Sigma_{\mu}$.
This is a necessary condition to obtain the ``self-dual'' gauge field strength 
on the $\mathbf{R}^8$. 
This condition corresponds to the extended ADHM equations in eight dimensions 
both for the commutative and the noncommutative case.

Before writing this condition, we notice that 
there are equivalence relations between different sets of 
$\textbf{A},\textbf{B}$ and $\textbf{B}'$,
\begin{equation}
  \mathbf{A}\sim U\mathbf{A}M\,,\quad \mathbf{B}\sim U\mathbf{B}M\,,
  \quad \mathbf{B}^{\prime}\sim U\mathbf{B}^{\prime}M\,,\label{eq:2.16}
\end{equation}
where $U\in U(N+2k)$ and $M\in GL(2k,\,\mathbf{C})$. 
The gauge field is invariant under this transformation. 
Using these relations, we can reduce $\textbf{A},\textbf{B}$ and $\textbf{B}'$
in the form:
\begin{equation}
\mathbf{A} = \left( \begin{array}{@{\,}cc@{\,}}
   A_2 & A_1\\
  -A_1^{\dagger}  &  A_2^{\dagger}\\
  I  &  J
  \end{array}  \right),\quad \mathbf{B}=\left( \begin{array}{@{\,}cc@{\,}}
   \textbf{1}_k & 0\\
    0    &  \textbf{1}_k    \\
  0  &  0
  \end{array}  \right),\quad 
\mathbf{B}' = \left( \begin{array}{@{\,}cc@{\,}}
   B_2 & B_1\\
  -B_1^{\dagger}  &  B_2^{\dagger}\\
  K  &  L
  \end{array}  \right)\,.\label{eq:2.17}
\end{equation}
Here $A_i$ and $B_i$ $(i=1,\,2)$ are $k\times k$ matrices, 
and $I,\,J,\,K$ and $L$ are $N\times k$ matrices.
In this representation, we can write down the condition 
that the $D^{\dagger}_zD_z$ commutes with $\Sigma_{\mu}$~\cite{ohta}.

In this way, the extended ADHM equations in the commutative case 
can be obtained as
\begin{equation}
\mu_{\mathbf{R}}^1=\mu_{\mathbf{C}}^1=\mu_{\mathbf{C}}^2 
 =\mu_{\mathbf{C}}^{2\,\,\,\prime}=\mu_{\mathbf{R}}^3  
=\mu_{\mathbf{C}}^3=0\,,\label{eq:2.18}
\end{equation}
where several quantities are defined by
\begin{eqnarray}
\mu_{\mathbf{R}}^1 &=& \left[ A_2^{\dagger},A_2\right] 
  -\left[ A_1^{\dagger},A_1\right]+I^{\dagger}I-J^{\dagger}J\,,\nonumber\\
\mu_{\mathbf{C}}^1 &=& \left[ A_2^{\dagger},A_1\right]+I^{\dagger}J\,,
                        \nonumber\\
\mu_{\mathbf{C}}^2 &=&\left[ A_2^{\dagger},B_2\right] 
  -\left[ B_1^{\dagger},A_1\right]+I^{\dagger}K-L^{\dagger}J\,,
\label{eq:2.19}\\
\mu_{\mathbf{C}}^{2\,\,\,\prime}  &=& \left[ A_2^{\dagger},B_1\right] 
  +\left[ B_2^{\dagger},A_1\right]+I^{\dagger}L+K^{\dagger}J\,,\nonumber\\
\mu_{\mathbf{R}}^3 &=& \left[ B_2^{\dagger},B_2\right] 
  -\left[ B_1^{\dagger},B_1\right]+K^{\dagger}K-L^{\dagger}L\,,\nonumber\\
\mu_{\mathbf{C}}^3 &=& \left[ B_2^{\dagger},B_1\right]+K^{\dagger}L\,.\nonumber
\end{eqnarray}
There are two real and four complex equations, 
which are related to the adjoint representation $\textbf{10}$ of $Sp(2)$.
Therefore the moduli space of the eight dimensional instantons 
we consider here is expected to have 
the structure of the $Sp(2)$ holonomy.
Examples of the $Sp(2)$ holonomy manifold are given in Ref.~\cite{ggpt, cglp}.


\section{Yang-Mills Theory on noncommutative space}


In this section, we briefly review the Yang-Mills theory 
on the noncommutative space. 
A field theory on the noncommutative space is defined 
by deforming the ring of functions on it. 
More concretely, the product of functions $f$ and $g$ is replaced 
with the Moyal star product that is defined by
\begin{equation}
(f\star g) (x)\equiv e^{\frac{i}{2} \theta^{\mu\nu} 
\partial_{\mu}\partial_{\nu}^{\prime}}f(x)g(x')|_{x=x'}\,.\label{eq:3.1}
\end{equation}
This equation implies that
\begin{equation}
 \left[ x^{\mu},\,x^{\nu}\right] =x^{\mu}\star x^{\nu}-x^{\nu}\star x^{\mu}
=i\,\theta^{\mu\nu}\,.\label{eq:3.2}
\end{equation}
This commutation relation characterizes the noncommutative space 
we treat in this paper.
The constant $\theta^{\mu\nu}$ is called the noncommutativity parameter.

There is another equivalent description of the noncommutative space, 
which is called the operator formalism and 
is useful for explicit calculations. 
These two descriptions are related via the Moyal-Weyl correspondence. 
In the operator formalism, we regard the coordinates as operators.
In this section, we denote the hat on the operators 
in order to emphasize that they are operators. 
The commutation relation between coordinates becomes as follows
\begin{equation}
\left[ \hat{x}^{\mu},\,\hat{x}^{\nu}\right]=i\,\theta^{\mu\nu}\,,
\label{eq:3.3}
\end{equation}
This relation is represented by using operators 
which act on the Hilbert space $\mathcal{H}$.

The derivative of an operator $\mathcal{O}$ is defined by
\begin{equation}
\partial_{\mu}\mathcal{O}\equiv \left[ \hat{\partial}_{\mu},\,\mathcal{O}
\right]\quad \textrm{where}\quad 
\hat{\partial}_{\mu}\equiv -i\,(\theta^{-1})_{\mu\nu}\hat{x}^{\nu}\,.
\label{eq:3.4}
\end{equation}
This derivative satisfies the Leibniz rule and the relations 
\begin{equation}
\partial_{\mu}\hat{x}^{\nu}=\delta_{\mu}{}^{\nu}\quad 
\textrm{and}\quad 
\left[ \hat{\partial}_{\mu},\,\hat{\partial}_{\nu}\right]=
i\,(\theta^{-1})_{\mu\nu}\,.\label{eq:3.5}
\end{equation}
The integral of an operator $\mathcal{O}$ is defined 
by the trace over the Hilbert space $\mathcal{H}$ as follows
\begin{equation}
   \int d^Dx \,\mathcal{O}(x)\equiv (2\pi)^D \sqrt{\mathrm{det}\,\theta}\,
 \mathrm{Tr}_{\mathcal{H}}\mathcal{O}(x)\,.\label{eq:3.6}
\end{equation}

We note that the strength of the gauge field $\hat{A}_{\mu}$ can be written as
\begin{equation}
\hat{F}_{\mu\nu} = \left[ \hat{X}_{\mu},\,\hat{X}_{\nu}\right]-i\,(\theta^{-1})
_{\mu\nu}\,,\label{eq:3.7}
\end{equation}
where the anti-hermitian operator $\hat{X}_{\mu}$ is defined by
\begin{equation}
 \hat{X}_{\mu}\equiv \hat{\partial}_{\mu}+\hat{A}_{\mu}\,.\label{eq:3.8}
\end{equation}
In this way the action of the noncommutative Yang-Mills theory
is expressed by
\begin{equation}
 S=-\frac{1}{4}(2\pi)^D \sqrt{\mathrm{det}\,\theta}\,
 \mathrm{Tr}_{\mathcal{H}} \mathrm{tr}_{U(N)}
 \hat{F}_{\mu\nu}\hat{F}_{\mu\nu}\,.\label{eq:3.9}
\end{equation}
Here $\mathrm{tr}_{U(N)}$ denotes the trace over the $U(N)$ matrix. 
In the following sections, we omit the hat on operators 
for simplicity of the description.


\section{$U(1)$ instanton in the case of anti-self-dual noncommutativity}


In this section, 
we study the noncommutative $U(1)$ instantons on $\mathbf{R}^8$ 
in the case that the noncommutativity parameter is anti-self-dual. 
In a wide subclass of the extended ADHM data, 
we show that the integer $k$ which appears in the extended ADHM construction 
should be interpreted as the $D4$-brane charge 
rather than the $D0$-brane charge.

As in the four dimensional case, 
it is easy to generalize the extended ADHM construction 
in eight dimensions to the noncommutative space 
because of its algebraic nature.
Since we define the instantons as ``self-dual'' configurations, 
the ``anti-self-dual'' noncommutativity parameter 
is expected to be of interest 
from the viewpoint of the resolution of the instanton moduli space.

Concretely, we introduce the anti-self-dual noncommutativity parameter 
as follows
\begin{equation}
  \theta^{13}=-\theta^{57}=\theta^{24}=-\theta^{68}=\frac{\zeta}{4} 
  \quad (\zeta >0)\,.
\label{eq:4.0.1}
\end{equation}
This implies the commutation relations of the complex coordinates:
\begin{eqnarray}
\left[ z_1,\,\bar{z}_1\right]= -\left[ z_2,\,\bar{z}_2\right]= 
\left[ z_3,\,\bar{z}_3\right]= -\left[ z_4,\,\bar{z}_4\right]=
-\frac{\zeta}{2}\,,\quad \textrm{others are zero}\,. \label{eq:4.0.2}
\end{eqnarray}
These relations are the same as those of the harmonic oscillators 
up to the multiplication of constants. 
Therefore we define the creation and annihilation operators by
\begin{equation}
 a^{\dagger}_m=\sqrt{\frac{2}{\zeta}}z_m\,,\quad 
 a_m=\sqrt{\frac{2}{\zeta}}\bar{z}_m\quad
\textrm{for}\quad m=1,\,3\,,\label{eq:4.0.3}
\end{equation}
as well as
\begin{equation} 
a^{\dagger}_m=\sqrt{\frac{2}{\zeta}}\bar{z}_m\,,\quad 
 a_m=\sqrt{\frac{2}{\zeta}} z_m\quad
\textrm{for}\quad m=2,\,4\,.\label{eq:4.0.4}
\end{equation}
The number operators can also be defined as
\begin{equation}
 n_m=a_m^{\dagger}a_m=
\left\{ 
\begin{array}{@{\,}lll}
\frac{2}{\zeta} z_m\bar{z}_m  & \textrm{for} &  m=1,\,3\,,\\
\frac{2}{\zeta} \bar{z}_mz_m  & \textrm{for} &  m=2,\,4\,.
\end{array}
\right.\label{eq:4.0.5}
\end{equation}
The Fock space $\mathcal{H}$ on which the creation and annihilation operators 
(\ref{eq:4.0.3}) and (\ref{eq:4.0.4}) act 
is spanned by the direct product of the Fock state: 
$|n_1:n_2:n_3:n_4\rangle \equiv |n_1\rangle \otimes |n_2\rangle \otimes 
 |n_3\rangle \otimes |n_4 \rangle$.
The creation and annihilation operators act on each Fock state as follows
\begin{equation}
  a_m|n_m\rangle =\sqrt{n_m}|n_m-1\rangle\,,\quad 
  a_m^{\dagger}|n_m\rangle =\sqrt{n_{m}+1}|n_m+1\rangle\quad
 \textrm{for}\quad m=1,\,\cdots,\,4\,. \label{eq:4.0.6}
\end{equation}

The noncommutativity of the complex coordinates (\ref{eq:4.0.2}) deforms 
the extended ADHM equations (\ref{eq:2.18}) as follows
\begin{equation}
 \mu_{\mathbf{R}}^1=\zeta (\mathbf{1}_{k\times k}+\Xi)\,,\quad 
 \mu_{\mathbf{C}}^1=\mu_{\mathbf{C}}^2 =\mu_{\mathbf{C}}^{2\,\,\,\prime}
 =\mu_{\mathbf{R}}^3 =\mu_{\mathbf{C}}^3=0\,, \label{eq:4.0.7}
\end{equation}
where $\Xi$ is defined by
\begin{equation}
 \Xi \equiv \frac{1}{2}\left( \{ B_2^{\dagger},\,B_2\} +
\{ B_1^{\dagger},\,B_1\} +K^{\dagger}K +L^{\dagger}L\right)\,.
\label{eq:4.0.8}
\end{equation}
These equations were originally obtained in Ref.~\cite{ohta}.


\subsection{$U(1)$, $k=1$ solution}


In this subsection, 
we start by constructing the $U(1)$, $k=1$ solution explicitly 
and discuss its $D$-brane interpretation 
by calculating the topological charges and the value of the action. 
It becomes clear that the $U(1)$, $k=1$ solution 
should be interpreted as the bound state of the $D4$-brane 
and the $D8$-brane with the $B$-field
rather than that of the $D0$-brane and the $D8$-brane with the $B$-field. 

An important fact for the $U(1)$ case is that 
we are allowed to take $J=K=L=0$~\cite{furuuchi} . 
Then we can obtain the non-trivial solution of the noncommutative version 
of the extended ADHM equations (\ref{eq:4.0.7}) by
\begin{equation}
 A_1=A_2=B_1=0\,,\quad B_2=a\,,\quad I=\sqrt{\zeta (1+a^2)}\,,
\label{eq:4.1.1}
\end{equation}
where the parameter $a$ is an arbitrary real number.
The Dirac-like operator becomes as follows
\begin{equation}
 D_z^{\dagger} = 
\left( \begin{array}{@{\,}ccc@{\,}}
   \bar{z}_2+a\bar{z}_4 & -z_1-az_3  &  \sqrt{\zeta (1+a^2)} \\
   \bar{z}_1+a\bar{z}_3  & z_2+az_4  &  0 
  \end{array}  \right)\,.\label{eq:4.1.2}
\end{equation}
The zero mode $\psi$ of $D^{\dagger}_z$ is a $3\times 1$ matrix
which is written as $ \psi \equiv 
\left( \begin{array}{@{\,}ccc@{\,}}
   \psi_1 & \psi_2 & \xi   
  \end{array}  \right)^{T}$\,.
Each component of $\psi$ is explicitly calculated as
\begin{eqnarray}
\psi_1 &=& -\frac{I}{\delta +a\eta +I^2/2}
\frac{1}{\sqrt{ 1+I^2 (\delta +a\eta +I^2/2)^{-1}}}
\,(z_2+az_4)\,,\nonumber\\
\psi_2 &=& \frac{I}{\delta +a\eta +I^2/2}
\frac{1}{\sqrt{ 1+I^2 (\delta +a\eta +I^2/2 )^{-1}}}
\,(\bar{z}_1+a\bar{z}_3)\,,\label{eq:4.1.3}\\
\xi &=& \frac{1}{\sqrt{ 1+I^2 (\delta +a\eta)^{-1}}}\,,\nonumber
\end{eqnarray}
where we defined the quantities $\delta$ and $\eta$ by
\begin{eqnarray} 
& & \delta \equiv 
z_1\bar{z}_1 +\bar{z}_2z_2 +a^2(z_3\bar{z}_3 +\bar{z}_4z_4) \,, 
  \label{eq:4.1.4}\\
& & \eta \equiv \bar{z}_4z_2+\bar{z}_2z_4 +z_1\bar{z}_3+z_3\bar{z}_1\,.
  \label{eq:4.1.5}
\end{eqnarray}
The following formulae are useful for the calculations, 
\begin{eqnarray}
& & (\bar{z}_1 +a\bar{z}_3)\,f(\delta +a\eta) 
=f(\delta +a\eta +I^2/2 )
(\bar{z}_1+a\bar{z}_3)\,,\label{eq:4.1.6}\\
& & (\bar{z}_2 +a\bar{z}_4)\,f(\delta +a\eta ) 
=f(\delta +a\eta -I^2 /2)
(\bar{z}_2+a\bar{z}_4)\,.\nonumber
\end{eqnarray}
Substituting the zero mode $\psi$ into the equations (\ref{eq:2.10}), 
the field strength form is explicitly obtained as follows
\begin{eqnarray}
F&=& F_{z_1\bar{z}_1}\bigl[ dz_1\wedge d\bar{z}_1 +dz_2\wedge d\bar{z}_2
+ a^2(dz_3\wedge d\bar{z}_3 +dz_4\wedge d\bar{z}_4) \bigr.\nonumber\\
& & \quad \quad \bigl.  +a(dz_1\wedge d\bar{z}_3 +dz_2\wedge d\bar{z}_4 
+dz_3\wedge d\bar{z}_1 +dz_4\wedge d\bar{z}_2) \bigr]\nonumber\\
& & \quad  +F_{z_1z_2}\left[ dz_1\wedge dz_2+ 
a(dz_1\wedge dz_4 +dz_3\wedge dz_2)
+ a^2\,dz_3\wedge dz_4 \right]\nonumber\\
& & \quad +F_{\bar{z}_1 \bar{z}_2}
\left[  d\bar{z}_1\wedge d\bar{z}_2 
+a( d\bar{z}_1\wedge d\bar{z}_4 +d\bar{z}_3\wedge d\bar{z}_2)
+a^2\,d\bar{z}_3\wedge d\bar{z}_4\right]\,,\label{eq:4.1.7}
\end{eqnarray} 
where the nontrivial components are explicitly given by
\begin{eqnarray}
F_{z_1\bar{z}_1} &=&  \frac{I^4}{(\delta +a\eta)(\delta +a\eta +I^2/2)
(\delta +a\eta +I^2)} \,J_3 \,,\nonumber\\
F_{z_1z_2} &=&  -\frac{\sqrt{2} I^4}{(\delta +a\eta)(\delta +a\eta +I^2/2)
(\delta +a\eta +I^2)} \,J_{+}\,,\label{eq:4.1.8}\\ 
F_{\bar{z}_1\bar{z}_2} &=& \frac{\sqrt{2} I^4}
{(\delta +a\eta)(\delta +a\eta +I^2/2)(\delta +a\eta +I^2)} 
\,J_{-} \,.\nonumber
\end{eqnarray}
Here as in the case of the four dimensional noncommutative instantons, 
we introduced the operators $J_{+},\,J_{-}$ and $J_3$~\cite{kly} by
\begin{eqnarray}
J_{+} &=& \frac{\sqrt{2}}{I^2}(\bar{z}_2+a\bar{z}_4)
(\bar{z}_1+a\bar{z}_3)\,,\nonumber\\
J_{-} &=& \frac{\sqrt{2}}{I^2}(z_1+az_3)(z_2+az_4)\,,\label{eq:4.1.9}\\
J_{3} &=& \frac{1}{I^2}\biggl[ (\bar{z}_2+a\bar{z}_4)(z_2+az_4)-
(z_1+az_3)(\bar{z}_1+a\bar{z}_3)\biggr]\,.\nonumber
\end{eqnarray}
These operators are found to satisfy the Lie algebra of $SU(2)$:
\begin{equation}
  \left[ J_{+},\,J_{-}\right] =J_3\,,\quad
  \left[ J_3,\,J_{\pm}\right] =\pm J_{\pm}\,.\label{eq:4.1.10}
\end{equation}

In the rest of this subsection, 
we discuss the properties of the above solution (\ref{eq:4.1.7}). 
At first, we can explicitly calculate the eight form charge $Q^{(8)}$:
\begin{equation}
 Q^{(8)} \equiv \frac{1}{4!(2\pi)^4}\int_{\mathbf{R}^8} 
 \,F\wedge F\wedge F\wedge F =0\,.
 \label{eq:4.1.11}
\end{equation}
Therefore the solution (\ref{eq:4.1.7}) does not have the $D0$-brane charge.

We can also calculate the value of the action of the solution as
\begin{eqnarray}
S &=&  -\frac{2I^4}{\zeta^2} \int d^8x \,
   \left( F_{z_1z_2}F_{\bar{z}_1\bar{z}_2}+ 
F_{\bar{z}_1\bar{z}_2}F_{z_1z_2}-2F_{z_1\bar{z}_1}F_{z_1\bar{z}_1}\right)
\nonumber\\
&=&  \pi^2 I^8 \left( \frac{\pi \zeta}{2}\right)^2 \mathrm{Tr}_{\mathcal{H}}
\frac{1}{(\delta +a\eta)(\delta +a\eta +I^2/2)^2(\delta +\eta +I^2)}\,.
\label{eq:4.1.12}
\end{eqnarray}
Here we used the integral formula (\ref{eq:3.6}) and the identity:
\begin{equation}
J_{+}J_{-}+J_{-}J_{+}+J_3^2 =\frac{1}{I^4}
 (\delta +a\eta)(\delta +a\eta +I^2)\,.\label{eq:4.1.13}
\end{equation}
To carry out the trace over the Hilbert space $\mathcal{H}$, 
we have to make the quantity $\delta +a\eta$ diagonal.
So we make the unitary transformation 
and define new creation and annihilation operators:
\begin{eqnarray}
& & \tilde{a}_1 =\frac{1}{\sqrt{1+a^2}}(a_1+aa_3)\,,\quad
 \tilde{a}_2 = \frac{1}{\sqrt{1+a^2}}(a_2+aa_4)\,,\nonumber\\
& & \tilde{a}_3 = \frac{1}{\sqrt{1+1/a^2}} 
\left(a_1-a^{-1}a_3\right)\,,\quad
 \tilde{a}_{4} = \frac{1}{\sqrt{1+1/a^2}}
\left(a_2-a^{-1}a_4\right)\,.\label{eq:4.1.14}
\end{eqnarray}
These new operators also satisfy the commutation relations 
of the harmonic oscillators,
\begin{equation}
\left[ \tilde{a}_{m},\,\tilde{a}_{n}^{\dagger}\right] = \delta_{m,n}\quad 
\textrm{for}\quad m,\,n =1,\,\cdots,\,4\,,\quad 
\textrm{others are zero},\label{eq:4.1.15}
\end{equation}
and the quantity $\delta +a\eta$ is made diagonal 
in the number basis of new harmonic oscillators (\ref{eq:4.1.14}) as
\begin{equation}
\delta +a\eta = 
\frac{I^2}{2} (\tilde{a}_{1}^{\dagger}\tilde{a}_{1} 
 +\tilde{a}_{2}^{\dagger}\tilde{a}_{2}) 
\equiv \frac{I^2}{2} (\tilde{n}_{1}+\tilde{n}_{2})\,.
\label{eq:4.1.16}
\end{equation}
Then we are able to carry out the calculation as follows
\begin{eqnarray}
S &=&  16\pi^2 \left( \frac{\pi\zeta}{2}\right)^2
 \sum_{(\tilde{n}_1,\,\tilde{n}_2,\,\tilde{n}_3,\,\tilde{n}_4) 
 \neq (0,\,0,\,0,\,0)}^{\infty}
\frac{1}{(\tilde{n}_1+\tilde{n}_2)(\tilde{n}_1+\tilde{n}_2+1)^2 
  ( \tilde{n}_1+\tilde{n}_2+2)}\nonumber\\
&=&  16\pi^2 \left( \frac{\pi\zeta}{2}\right)^2  
\sum_{(\tilde{n}_3,\,\tilde{n}_4)}^{\infty} \sum_{N=1}^{\infty}
\frac{1}{N(N+1)(N+2)}\nonumber\\
&=&   4\pi^2 \left( \frac{\pi\zeta}{2}\right)^2 
\sum_{(\tilde{n}_3,\,\tilde{n}_4)}^{\infty}=4\pi^2 V_4\,,\label{eq:4.1.17}
\end{eqnarray}
where we used the formula of the summation:
\begin{equation}
 \sum_{(\tilde{n}_1,\tilde{n}_2)\neq (0,0)}^{\infty} 
   \langle N|\mathcal{O}(\tilde{n}_1+\tilde{n}_2)|N
\rangle =\sum_{N=1}^{\infty}(N+1)
\langle N|\mathcal{O}(N)|N\rangle\,,\label{eq:4.1.18}
\end{equation}
and the formula of the four dimensional volume $V_4$ in the operator formalism:
\begin{equation}
V_4 \equiv \int d^4x =\left( \frac{\pi\zeta}{2}\right)^2 
 \sum_{(\tilde{n}_3,\,\tilde{n}_4)}^{\infty}\,.\label{eq:4.1.19}
\end{equation}
The appearance of the four dimensional volume $V_4$ 
suggests that the solution (\ref{eq:4.1.7}) 
has the four dimensional nature. 
Therefore it is natural to interpret the solution (\ref{eq:4.1.7})  
as the noncommutative version of the four dimensional instantons.

This four dimensional nature can be seen more explicitly 
by transforming the solution. 
Now let's define new coordinates $\tilde{z}_{m}$ $(m=1,\,\cdots,\,4)$ by  
\begin{eqnarray}
& & \tilde{z}_1=\frac{1}{\sqrt{1+a^2}}(z_1+az_3)\,,\quad 
    \tilde{z}_2=\frac{1}{\sqrt{1+a^2}}(z_2+az_4)\,,\label{eq:4.1.20}\\
& & \tilde{z}_3=\frac{1}{\sqrt{1+1/a^2}}(z_1-a^{-1}z_3)\,,\quad 
    \tilde{z}_4=\frac{1}{\sqrt{1+1/a^2}}(z_2-a^{-1}z_4)\,.\nonumber 
\end{eqnarray}
These new coordinates also satisfy the same commutation relations 
as (\ref{eq:4.0.2}):
 \begin{eqnarray}
\left[ \tilde{z}_1,\,\bar{\tilde{z}}_1\right] 
= -\left[ \tilde{z}_2,\,\bar{\tilde{z}}_2\right] 
= \left[ \tilde{z}_3,\,\bar{\tilde{z}}_3\right] 
= -\left[ \tilde{z}_4,\,\bar{\tilde{z}}_4\right] 
= -\frac{\zeta}{2}\,,\quad \textrm{others are zero}\,. 
\label{eq:4.0.20.1}
\end{eqnarray}
Then the solution (\ref{eq:4.1.7}) can be rewritten 
in these new coordinates as
\begin{equation}
F=\frac{\zeta^2}{\Delta (\Delta +\zeta /2)
(\Delta  +\zeta)} \,\left[ \tilde{J}_3\,
( d\tilde{z}_1\wedge d\bar{\tilde{z}}_1 
+d\tilde{z}_2\wedge d\bar{\tilde{z}}_2)
 -\sqrt{2} \tilde{J}_{+}\, d\tilde{z}_1\wedge d\tilde{z}_2 
+\sqrt{2}\tilde{J}_{-}\, 
 d\bar{\tilde{z}}_1\wedge d\bar{\tilde{z}}_2 \right]\,,\label{eq:4.1.21}
\end{equation} 
where we defined the quantity 
$\Delta \equiv \tilde{z}_1\bar{\tilde{z}}_{1} 
+\bar{\tilde{z}}_{2}\tilde{z}_2$ 
and the following operators: 
\begin{equation}
\tilde{J}_{+} = \frac{\sqrt{2}}{\zeta}\,
\bar{\tilde{z}}_2\bar{\tilde{z}}_1\,,\quad
\tilde{J}_{-} = \frac{\sqrt{2}}{\zeta}\,\tilde{z}_1\tilde{z}_2\,,\quad
\tilde{J}_{3} = \frac{1}{\zeta}\left( \bar{\tilde{z}}_2\tilde{z}_2 
- \tilde{z}_1\bar{\tilde{z}}_1\right)\,.\label{eq:4.1.22}
\end{equation}
These operators satisfy the Lie algebra (\ref{eq:4.1.10}) of $SU(2)$.
The above expression (\ref{eq:4.1.21}) is surely 
the $U(1)$ one instanton solution 
on the four dimensional space $\tilde{\mathbf{R}}^4$ spanned 
by the coordinates: 
$\tilde{z}_1$, $\bar{\tilde{z}}_{1}$, $\tilde{z}_2$ and $\bar{\tilde{z}}_{2}$. 
So the solution (\ref{eq:4.1.21}) has the four form charge
over the four dimensional subspace $\tilde{\mathbf{R}}^4$ as
\begin{equation}
Q^{(4)} \equiv -\frac{1}{2(2\pi)^2}\int_{\tilde{\mathbf{R}}^4} F\wedge F 
= 4 \sum_{N=1}^{\infty} \frac{1}{N(N +1)(N+2)}=+1\,.\label{eq:4.1.23}
\end{equation}

From the results (\ref{eq:4.1.11}), (\ref{eq:4.1.17}) and (\ref{eq:4.1.23}), 
the solution (\ref{eq:4.1.21}) should be interpreted 
as the bound state of the $D4$-brane and the $D8$-brane with the $B$ field, 
and we can naturally interpret the total value of the action (\ref{eq:4.1.17}) 
as the product of the action 
of the four dimensional instanton over $\tilde{\mathbf{R}}^4$ 
and the volume of the four dimensional space spanned by the coordinates: 
$\tilde{z}_3$, $\bar{\tilde{z}}_{3}$, $\tilde{z}_4$ and $\bar{\tilde{z}}_{4}$.

Here we comment on the problem associated with the zero mode, 
which we encounter when we normalize the $\psi$. 
If we rewrite the Dirac-like operator (\ref{eq:4.1.2}) 
by new coordinates (\ref{eq:4.1.20}), 
then the extended ADHM construction reduces to the ADHM construction 
of the noncommutative instanton 
in the four dimensional subspace $\tilde{\mathbf{R}}^4$. 
Therefore this problem of the zero mode is essentially the same 
as that of the noncommutative instanton
in $\tilde{\mathbf{R}}^4$~\cite{furuuchi}. 
We have used this fact and the procedure of Ref.~\cite{kly} 
when we calculate the value of the action (\ref{eq:4.1.17}) 
and the four form charge (\ref{eq:4.1.23}).


\subsection{Generalization to $U(1)$, multi $k$ case}


An important point in the previous subsection is that 
we are able to reduce the extended ADHM construction to 
the four dimensional ADHM construction 
by the unitary transformation of the coordinates. 
In this subsection, we comment on the generalization 
of this scenario to the more general case.

It is difficult to solve the noncommutative version 
of the extended ADHM equations (\ref{eq:4.0.7}) generally. 
However there is an interesting subspace 
in the moduli space of the extended ADHM data. 
If we take 
\begin{equation}
  B_1=w_1\mathbf{1}_{k\times k}\,,\quad 
  B_2=w_2\mathbf{1}_{k\times k}\,,\quad K=L=\mathbf{0}_{1\times k}\,,
  \label{eq:4.2.1}
\end{equation}
where $w_{i}\, (i=1,\,2)$ are arbitrary complex parameters, 
then the extended ADHM equations reduce to the equations 
similar to the four dimensional deformed ADHM equations:
\begin{equation}
  \mu^{1}_{\mathbf{R}}=\zeta (1+|w_1|^2 +|w_2|^2)\mathbf{1}_{k\times k}\,,
\quad \mu^1_{\mathbf{C}}=\mathbf{0}_{k\times k}\,.\label{eq:4.2.2}
\end{equation}
Here $\mathbf{0}_{1\times k}$ and $\mathbf{0}_{k\times k}$ 
denote respectively $1\times k$ and $k\times k$ matrices 
whose all components are zero, 
and $\mathbf{1}_{k\times k}$ denotes a $k\times k$ unit matrix. 
It is an easy problem to solve these reduced equations (\ref{eq:4.2.2}) 
since the solutions of the deformed ADHM equations in four dimensional case  
are well-known. 
Some relevant references are~\cite{
ns, nekrasov, furuuchi, furuuchi2, furuuchi3, 
agms, kly, ckt, hamanaka, popov2, sako1, sako2, sako3, 
meron, dress, ly, lty, kly2, ly2}.

We can generalize the procedure of the previous subsection 
to the present case. 
We are able to extend the definition (\ref{eq:4.1.20}) of new coordinates
to the case (\ref{eq:4.2.1}) straightforwardly. 
Then the extended Dirac-like operator (\ref{eq:2.7}) 
reduces to that of the ADHM construction of the instantons 
over the four dimensional subspace $\tilde{\mathbf{R}}^4$. 
Therefore we can naturally interpret the solution (\ref{eq:4.2.1}) 
as the bound state of the $k$ $D4$-branes and the $D8$-brane 
with the $B$-field, 
and we can show that the integer $k$ which appears 
in the extended ADHM construction 
should be interpreted as the $D4$-brane charge 
rather than the $D0$-brane charge 
when the noncommutativity parameter is anti-self-dual.


\section{Relationship with Solution Generating Technique}


In this section, 
we consider the case that the noncommutativity parameter is self-dual. 
We find the relationship with the solution generating technique 
and show that the integer $k$ can be interpreted 
as the charge of the $D0$-brane bound to the $D8$-brane with the $B$-field. 
In four dimensions, this relationship was established 
by Hamanaka~\cite{hamanaka}. 
However in the eight dimensions this relationship has not yet been found. 
We construct the solution of the extended ADHM equations, 
which corresponds to the localized instanton solution 
obtained by using the solution generating technique, 
and interpret it as the system of $k$ $D0$-branes 
and the $D8$-brane with the $B$-field.

We introduce the self-dual noncommutativity parameter as
\begin{equation}
 \theta^{13}=\theta^{57}=\theta^{24}=\theta^{68}=\frac{\zeta}{4} 
\quad (\zeta >0) \,.\label{eq:5.1}
\end{equation}
This implies the following commutation relations of the complex coordinates: 
\begin{equation}
\left[ z_1,\,\bar{z}_1\right]=\left[ z_2,\,\bar{z}_2\right]=
\left[ z_3,\,\bar{z}_3\right]=\left[ z_4,\,\bar{z}_4\right]= 
-\frac{\zeta}{2}\,,\quad 
\textrm{others are zero}.\label{eq:5.2}
\end{equation}
These relations are the same as those of the harmonic oscillators 
up to the multiplication of constants. 
Therefore we define the creation and annihilation operators by
\begin{equation}
 a^{\dagger}_m=\sqrt{\frac{2}{\zeta}}z_m\,,\quad 
 a_m=\sqrt{\frac{2}{\zeta}}\bar{z}_m\quad 
\textrm{for}\quad m=1,\,\cdots,\,4\,.\label{eq:5.3}
\end{equation}
The number operators can also be defined as
\begin{equation}
n_m=a_m^{\dagger}a_m=\frac{2}{\zeta} z_m\bar{z}_m\quad 
\textrm{for}\quad m=1,\,\cdots,\,4\,.\label{eq:5.4}
\end{equation}
As in the previous section, 
the Fock space $\mathcal{H}$ on which the creation and annihilation operators 
(\ref{eq:5.3}) act 
is spanned by the direct product of the Fock state: 
$|n_1:n_2:n_3:n_4\rangle \equiv |n_1\rangle \otimes |n_2\rangle \otimes 
 |n_3\rangle \otimes |n_4 \rangle$.
The creation and annihilation operators act on each Fock state 
in the same way as (\ref{eq:4.0.6}).

It is easily found that
the extended ADHM equations (\ref{eq:2.18}) in the commutative case 
are not deformed by the noncommutativity of the coordinates (\ref{eq:5.2}), 
\begin{equation}
\mu_{\mathbf{R}}^1=\mu_{\mathbf{C}}^1=\mu_{\mathbf{C}}^2 =\mu_{\mathbf{C}}^{2\,\,\,\prime}=\mu_{\mathbf{R}}^3 =\mu_{\mathbf{C}}^3=0\,.\label{eq:5.5}
\end{equation}

Now let's find the solution of the extended ADHM equations (\ref{eq:5.5}), 
which is related to the localized instanton solution 
obtained by using the solution generating technique. 
We consider the case of the gauge group $U(1)$ and multi $k$. 
It is allowed to take $J=K=L=\mathbf{0}_{1\times k}$ for the $U(1)$ case. 
Then the extended ADHM equations (\ref{eq:5.5}) are simply solved 
and the solution which might correspond to 
the localized instanton solution is obtained by
\begin{equation}
 A_1=A_2=B_1=\mathbf{0}_{k\times k}\,,\quad I=\mathbf{0}_{1\times k}\,,
\quad B_2=\mathbf{1}_{k\times k}\,.\label{eq:5.6}
\end{equation}

The extended ADHM construction gives rise to the instantons 
through the zero mode of the Dirac-like operator. 
So we need to look for the zero mode of the Dirac-like operator: 
\begin{equation}
D_z^{\dagger} =
\left( \begin{array}{@{\,}ccc@{\,}}
   (\bar{z}_2+\bar{z}_4) \mathbf{1}_{k\times k} & 
   -(z_1+z_3)\mathbf{1}_{k\times k} & 
   \mathbf{0}_{k\times 1} \\
   (\bar{z}_1+\bar{z}_3)\mathbf{1}_{k\times k}  & 
   (z_2+z_4) \mathbf{1}_{k\times k} &  
   \mathbf{0}_{k\times 1} 
  \end{array}  \right)\,.\label{eq:5.7}
\end{equation}
Seemingly the Dirac-like operator (\ref{eq:5.7}) has the only trivial solution.
We are however able to construct the non-trivial zero mode 
of the Dirac-like operator (\ref{eq:5.7}) 
by using the partial isometry in the noncommutative setting. 
Here, in order to write down the zero mode of $D_z^{\dagger}$, 
we prepare an ordering of the states:
\begin{equation}
|n_1:n_2:n_3:n_4\rangle =\prod_{i=1}^{4}\frac{1}{\sqrt{n_i!}} 
(a^{\dagger}_i)^{n_i}|0:0:0:0\rangle\,.\label{eq:5.8}
\end{equation}
Two sets of four non-negative integers, $\mathbf{m}=(m_1,\,\cdots,\,m_4)$ 
and $\mathbf{n}=(n_1,\,\cdots,\,n_4)$, 
for which we define $\bar{m}_j=\sum_{i=j}^{4}m_i$, and 
$\bar{n}_j=\sum_{i=j}^{4}n_i$ with $j=1,\,\cdots, 4$, 
are ordered by the following rules:
\begin{enumerate}
\item If $\bar{m}_j=\bar{n}_j$ for all $1\leq j\leq 4$,\, 
$\mathbf{m}=\mathbf{n}$\,.
\item If $\bar{m}_j=\bar{n}_j\, (j=1,\cdots, k-1)$ and 
$\bar{m}_k >\bar{n}_k$ for some $k \,(1\leq k \leq 4)$,\,
$\mathbf{m}> \mathbf{n}\,.$
 \item If $\bar{m}_j=\bar{n}_j\, (j=1,\cdots, k-1)$ and 
$\bar{m}_k < \bar{n}_k$ for some $k \,(1\leq k \leq 4)$,\,
$\mathbf{m} < \mathbf{n}\,.$
\end{enumerate}
We can order all the states by these rules. 
For example, these rules order the states as
\begin{eqnarray}
& & |0\rangle\!\rangle =|0:0:0:0\rangle\,,\nonumber\\
& & |1\rangle\!\rangle =|1:0:0:0\rangle\,,\quad 
    |2\rangle\!\rangle =|0:1:0:0\rangle\,,\quad
    |3\rangle\!\rangle =|0:0:1:0\rangle\,,\nonumber\\
& & |4\rangle\!\rangle =|0:0:0:1\rangle\,,\quad
    |5\rangle\!\rangle =|2:0:0:0\rangle\,,\quad \cdots\,.\label{eq:5.9}
\end{eqnarray}

The zero mode $\psi$ of $D^{\dagger}_z$ is a $(2k+1)\times 1$ matrix
which is written as $ \psi \equiv 
\left( \begin{array}{@{\,}ccc@{\,}}
   \psi_1 & \psi_2 & \xi   
  \end{array}  \right)^{T}$\,. 
Here $\psi_1$ and $\psi_2$ are $k\times 1$ matrices respectively, 
and $\xi$ is a $1\times 1$ matrix.   
Each component of $\psi$ is explicitly obtained as
\begin{equation}
 \psi_1 =
\left( \begin{array}{@{\,}c@{\,}}
    |0\rangle\!\rangle \langle\!\langle 0 |  \\
    |0\rangle\!\rangle \langle\!\langle 1 |  \\
      \vdots               \\
    |0\rangle\!\rangle \langle\!\langle k-1 |  \\
  \end{array}  \right) \,,
\quad \psi_2=
\left( \begin{array}{@{\,}c@{\,}}
    0  \\
    0  \\
      \vdots               \\
    0  \\
  \end{array}  \right)
\,,\quad   \xi =S_k\,,\label{eq:5.10}
\end{equation}
where we defined the shift operator:
\begin{equation}
 S_k\equiv \sum_{i=0}^{\infty} 
|i\rangle\!\rangle \langle\!\langle i+k |\,.\label{eq:5.11}
\end{equation}
This shift operator is the typical example of the partial isometry, 
which is used in the construction by using the solution generating technique, 
and satisfies
\begin{equation}
 S_kS_k^{\dagger}=1\,,\quad S_k^{\dagger}S_k=1-P_k\,.\label{eq:5.12}
\end{equation}
Here we defined the projection operator of rank $k$ by
\begin{equation}
P_k \equiv \sum_{i=0}^{k-1}|i\rangle\!\rangle \langle\!\langle i| \,,
\label{eq:5.13}
\end{equation}
which satisfies
\begin{equation}
 S_kP_k=P_kS_k^{\dagger}=0\,.\label{eq:5.14}
\end{equation}

Then we can easily write down the explicit expression of 
the gauge field (\ref{eq:3.8}):
\begin{eqnarray}
X_{\mu}
&=& \psi^{\dagger}\left[ \hat{\partial}_{\mu},\,\psi\right] 
 +\hat{\partial}_{\mu}
= \psi^{\dagger}\hat{\partial}_{\mu}\psi \nonumber\\
&=& S_k^{\dagger}\,\hat{\partial}_{\mu}S_k \,,
\label{eq:5.15}
\end{eqnarray}
and the strength of gauge field (\ref{eq:3.7}):
\begin{equation}
F_{\mu\nu}= i\,(\theta^{-1})_{\mu\nu}(S_kS_k^{\dagger}-1)= 
-i\,(\theta^{-1})_{\mu\nu}P_k\,.\label{eq:5.16}
\end{equation}
From the noncommutativity parameter (\ref{eq:5.1}) in this case, 
the field strength form is written as
\begin{equation}
F= \frac{2}{\zeta}P_k\,\left( dz_1\wedge d\bar{z}_1 +dz_2\wedge d\bar{z}_2
+ dz_3\wedge d\bar{z}_3 +dz_4\wedge d\bar{z}_4\right)\,.\label{eq:5.17}
\end{equation}
The solution (\ref{eq:5.15}) for $k=1$ was originally obtained 
in Ref.~\cite{fio} by using the solution generating technique, 
and it was confirmed that the solution (\ref{eq:5.15}) 
preserves 3/16 of supersymmetries 
by investigating small fluctuations around the solution. 
The condition for preserving 3/16 of supersymmetries, 
which was found in Ref.~\cite{fio}, 
corresponds to choose the noncommutativity parameter as (\ref{eq:5.1}). 
Then it is confirmed that the solution (\ref{eq:5.6}) 
of the extended ADHM equations 
corresponds to the solution constructed 
by using the solution generating technique.

In the rest of this section, we study the properties 
of the above solution (\ref{eq:5.15}). 
At first, we are able to calculate the eight form charge $Q^{(8)}$ as 
\begin{equation}
 Q^{(8)}\equiv \frac{1}{4!(2\pi)^4}\int_{\mathbf{R}^8} 
 \,F\wedge F\wedge F\wedge F =\textrm{Tr}_{\mathcal{H}}\,P_k = k\,,
\label{eq:5.18}
\end{equation}
where we used the formula:
\begin{equation}
 dz^1\wedge d\bar{z}^1 \wedge dz^2\wedge d\bar{z}^2
 \wedge dz^3\wedge d\bar{z}^3  \wedge dz^4\wedge d\bar{z}^4= 
16(\textrm{volume form})\,.\label{eq:5.19}
\end{equation}
Therefore the integer $k$ which appears in the extended ADHM construction 
can be regarded as the $D0$-brane charge. 
We can also calculate the value of the action for the  solution, 
\begin{eqnarray}
 S &=& -\frac{1}{2} \int d^8x\, 
        (F_{13}^2+F_{57}^2+F_{24}^2+F_{68}^2)\nonumber\\
   &=&  2\pi^4\zeta^2\, \textrm{Tr}_{\mathcal{H}}\,P_k 
    =   2\pi^4\zeta^2 k\,.\label{eq:5.20}
\end{eqnarray}
From the results (\ref{eq:5.18}) and (\ref{eq:5.20}), 
the solution can be interpreted as the system of 
the $k$ $D0$-branes and the $D8$-brane with the $B$ field.


\section{Conclusion}


In this paper, we have studied the noncommutative version 
of the extended ADHM construction 
in the eight dimensional $U(1)$ Yang-Mills theory. 
We have found that it is an important difference 
whether the noncommutativity parameter is anti-self-dual or self-dual. 
In the case that the noncommutativity parameter is anti-self-dual, 
we have shown that the integer $k$ 
which appears in the extended ADHM construction 
should be interpreted as the $D4$-brane charge 
rather than the $D0$-brane charge. 
We have confirmed this fact in a wide subspace of the extended ADHM data 
by calculating the topological charges.

We have found the relationship with the solution generating technique 
in the case that the noncommutativity parameter is self-dual. 
The shift operator of the solution generating technique 
has naturally appeared in the extended ADHM construction. 
We have also shown in this case that the integer $k$ can be interpreted 
as the charge of the $D0$-brane bound to the $D8$-brane with the $B$-field

The natural generalization of our study is 
to consider the gauge group of the higher rank 
since our study has been restricted to the case of the $U(1)$ gauge group. 
It should be checked whether the qualitative nature we have found 
changes or not. 
For example, it is of interest whether the integer $k$ 
which appears in the extended ADHM construction 
becomes to be interpreted as the $D0$-brane charge or not
in the case that the noncommutativity parameter is anti-self-dual.
Another generalization of our study is 
to construct the solutions of the BPS equations (\ref{eq:2.1}) 
except for the case related to the $Sp(2)$ symmetry 
in the noncommutative Yang-Mills theory.

The moduli space of the noncommutative instantons in eight dimensions 
is expected to have much richer structure 
than that of the noncommutative instantons in four dimensions. 
Therefore further investigation is necessary to understand the topological 
structure of gauge fields on the noncommutative $\mathbf{R}^8$.


\bigskip
\bigskip
\centerline{\bf Acknowledgments}

\vskip 0.6cm

We would like to thank S. Watamura 
for useful comments, reading manuscripts and encouragements.

\newpage


\bibliography{}

\begin{thebibliography}{999}

\bibitem{cds}
A. Connes, M. R. Douglas and A. Schwarz, 
JHEP \textbf{9802} (1998) 003, 
hep-th/9711162.

\bibitem{sw}
N. Seiberg and E. Witten, 
JHEP \textbf{9909} (1999) 032, 
hep-th/9908142.

\bibitem{ns}
N. Nekrasov and A. Schwarz,  
Comm. Math. Phys. \textbf{198} (1998) 689-703, 
hep-th/9802068.

\bibitem{nekrasov}
N. Nekrasov, 
``Trieste lectures on solitons in noncommutative gauge theories'', 
hep-th/0011095.

\bibitem{cdfn}
E. Corrigan, C. Devchand, D. B. Fairlie and J. Nuyts, 
Nucl. Phys. \textbf{B214} (1983) 452.

\bibitem{ward}
R. S. Ward, 
Nucl. Phys. \textbf{B239} (1984) 381-396.

\bibitem{cgk}
E. Corrigan, P. Goddard and A. Kent, 
Comm. Math. Phys. \textbf{100} (1985) 1-13.

\bibitem{hull}
C. M. Hull, 
Adv. Theor. Math. Phys. \textbf{2} (1998) 619-632, 
hep-th/9710165.

\bibitem{ohtan}
N. Ohta and P. K. Townsend, 
Phys. Lett. \textbf{B418} (1998) 77-84, 
hep-th/9710129.

\bibitem{cimm}
B. Chen, H. Itoyama, T. Matsuo and K. Murakami, 
Nucl. Phys. \textbf{B576} (2000) 177-195, 
hep-th/9910263.

\bibitem{park}
M. Mihailescu, I. Y. Park and T. A. Tran,  
Phys. Rev. \textbf{D64} (2001) 046006, 
hep-th/0011079.

\bibitem{witten}
E. Witten, 
``BPS Bound States of $D0$-$D6$ and $D0$-$D8$ Systems in a $B$-field'', 
JHEP \textbf{0204} (2002) 012, 
hep-th/0012054.

\bibitem{sato}
M. Sato, 
``BPS Bound States of D6-branes and Lower Dimensional D-branes'', 
Int. J. Mod. Phys. \textbf{A16} (2001) 4069, 
hep-th/0101226.

\bibitem{fio}
A. Fujii, Y. Imaizumi and N. Ohta, 
``Supersymmetry, Spectrum and Fate of D0-D$p$ Systems with $B$-field'', 
Nucl. Phys. \textbf{B615} (2001) 61-81, 
hep-th/0105079.

\bibitem{ohta}
K. Ohta, 
``Supersymmetric D-brane Bound States with $B$-field and
Higher Dimensional Instantons on Noncommutative Geometry'', 
Phys. Rev. \textbf{D64} (2001) 046003, 
hep-th/0101082.

\bibitem{hio}
M. Hamanaka, Y. Imaizumi and N. Ohta, 
``Moduli Space and Scattering of D0-branes 
in Noncommutative Super Yang-Mills Theory'', 
Phys. Lett. \textbf{B529} (2002) 163-170, 
hep-th/0112050.

\bibitem{kly3}
C. Kim, K. Lee and S.H. Yi,  
Phys. Lett. \textbf{B543} (2002) 107-114, 
hep-th/0204109.

\bibitem{pt}
G. Papadopoulos and A. Teschendorff, 
``Instantons at Angles'', 
Phys. Lett. \textbf{B419} (1998) 115-122, 
hep-th/9708116.

\bibitem{hiraoka}
Y. Hiraoka, 
``Eight Dimensional Noncommutative Instantons and 
D0-D8 Bound States with $B$-field'', 
Phys. Lett. \textbf{B536} (2002) 147-153, 
hep-th/0203047.

\bibitem{hiraoka2}
Y. Hiraoka, 
``BPS Solutions of Noncommutative Gauge Theories 
in Four and Eight Dimensions'', 
hep-th/0205010.

\bibitem{blp}
D. Bak, K. Lee and J. H. Park, 
``BPS Equations in Six and Eight Dimensions'', 
Phys. Rev. \textbf{D66} (2002) 025021, 
hep-th/0204221.

\bibitem{valtancoli}
P. Valtancoli, 
``Noncommutative Instantons on $d=2n$ Planes from Matrix Models'', 
hep-th/0209118.

\bibitem{adhm}
M. Atiyah, N. Hitchin, V. Drinfeld and Y. Manin, 
Phys. Lett. \textbf{65B} (1978) 185.

\bibitem{cg}
E. Corrigan and P. Goddard, 
Ann. Phys. \textbf{154} (1984) 253-279.

\bibitem{furuuchi}
K. Furuuchi, 
``Topological Charges of $U(1)$ Instantons'', 
Prog. Theor. Phys. Suppl. \textbf{144} (2001) 79-91, 
hep-th/0010006.

\bibitem{furuuchi2}
K. Furuuchi, 
``Instantons on Noncommutative $\mathbf{R}^4$ and Projection Operators'', 
Prog. Theor. Phys. \textbf{103} (2000) 1043-1068, 
hep-th/9912047.

\bibitem{furuuchi3}
K. Furuuchi, 
``Dp-D(p+4) in Noncommutative Yang-Mills'', 
JHEP \textbf{0103} (2001) 033, 
hep-th/0010119.

\bibitem{agms}
M. Aganagic, R. Gopakumar, S. Minwalla and A. Strominger, 
``Unstable Solitons in Noncommutative Gauge Theory'', 
JHEP \textbf{0104} (2001) 001, 
hep-th/0009142.

\bibitem{kly}
K. Kim, H. Lee and H. S. Yang, 
``Comments on Instantons on Noncommutative $\mathbf{R}^4$'', 
J. Korean Phys. Soc. \textbf{41} (2002) 290-297, 
hep-th/0003093.

\bibitem{ckt}
C. H. Chu, V. V. Khoze and G. Travaglini, 
``Notes on Noncommutative Instantons'', 
Nucl. Phys. \textbf{B621} (2002) 101-130, 
hep-th/0108007.

\bibitem{hamanaka}
M. Hamanaka, 
``ADHM/Nahm Construction of Localized Solitons in 
Noncommutative Gauge Theories'', 
Phys. Rev. \textbf{D65} (2002) 085022, 
hep-th/0109070.

\bibitem{popov2}
O. Lechtenfeld and A. D. Popov, 
``Noncommutative 't Hooft Instantons'', 
JHEP \textbf{0203} (2002) 040, 
hep-th/0109209.

\bibitem{sako1}
T. Ishikawa, S. Kuroki and A. Sako, 
``Elongated U(1) Instantons on Noncommutative $\mathbf{R}^4$'', 
JHEP \textbf{0111} (2001) 068, 
hep-th/0109111.

\bibitem{sako2}
T. Ishikawa, S. Kuroki and A. Sako, 
``Instanton Number Calculus on Noncommutative $\mathbf{R}^4$'', 
JHEP \textbf{0208} (2002) 028, 
hep-th/0201196.

\bibitem{sako3}
A. Sako, 
``Instanton Number of Noncommutative U(n) Gauge Theory'', 
hep-th/0209139.

\bibitem{meron}
F. Franco-Sollova and T. Ivanova, 
``On Noncommutative Merons and Instantons'', 
hep-th/0209153.

\bibitem{dress}
Z. Horv\'{a}th, O. Lechtenfeld and M. Wolf, 
``Noncommutative Instantons via Dressing and Splitting Approaches'', 
hep-th/0211041.

\bibitem{ly}
K. Lee and P. Yi, 
Phys. Rev. \textbf{D61} (2000) 125015, 
hep-th/9911186.

\bibitem{lty}
K. Lee, D. Tong and S. Yi,  
Phys. Rev. \textbf{D63} (2001) 065017, 
hep-th/0008092.

\bibitem{kly2}
K. Kim, B. Lee and H. Yang, 
Phys. Rev. \textbf{D66} (2002) 025034, 
hep-th/0205010.

\bibitem{ly2}
B. Lee and H. Yang,  
Phys. Rev. \textbf{D66} (2002) 045027, 
hep-th/0206001.

\bibitem{ggpt}
J. P. Gauntlett, G. W. Gibbons, G. Papadopoulos and P. K. Townsend, 
``Hyper-K\"{a}hler manifolds and multiply-intersecting branes'', 
Nucl. Phys. \textbf{B500} (1997) 133-162, 
hep-th/9702202.

\bibitem{cglp}
M. Cveti$\check{\mathrm{c}}$, G. W. Gibbons, H. L\"{u} and C. N. Pope, 
``Hyper-K\"{a}hler Calabi metrics, $L^2$ harmonic forms, 
resolved M2-branes, and AdS${}_4$/CFT${}_{3}$ correspondence'', 
hep-th/0102185.

\end{thebibliography}


\end{document}